\documentclass[english, aps,prd,twocolumn,twoside,superscriptaddress,floatfix, nofootinbib]{revtex4-1}

\usepackage[utf8]{inputenc}
\usepackage{amsmath}
\usepackage{geometry}
\usepackage{xcolor}
\usepackage{graphicx}
\usepackage{float}
\usepackage{lipsum, babel}
\usepackage{graphicx}
\usepackage{color}
\usepackage{hyperref}
\usepackage{ifthen,amsmath,amssymb, bm}

\newcommand{\vL} { {\bm{L}} }

\newcommand{\beq} {\begin{equation}}
\newcommand{\eeq} {\end{equation}}
\newcommand{\bal} {\begin{aligned}}
\newcommand{\eal} {\end{aligned}}

\begin{document}

\title{Optimal multifrequency weighting for CMB lensing}

\author{Noah Sailer}
\email{nsailer@berkeley.edu}
\affiliation{Berkeley Center for Cosmological Physics, Department of Physics,
University of California, Berkeley, CA 94720, USA}
\affiliation{Lawrence Berkeley National Laboratory, One Cyclotron Road, Berkeley, CA 94720, USA}
\author{Emmanuel Schaan}
\affiliation{Lawrence Berkeley National Laboratory, One Cyclotron Road, Berkeley, CA 94720, USA}
\affiliation{Berkeley Center for Cosmological Physics, Department of Physics,
University of California, Berkeley, CA 94720, USA}
\author{Simone Ferraro}
\affiliation{Lawrence Berkeley National Laboratory, One Cyclotron Road, Berkeley, CA 94720, USA}
\affiliation{Berkeley Center for Cosmological Physics, Department of Physics,
University of California, Berkeley, CA 94720, USA}
\author{Omar Darwish}
\affiliation{Department of Applied Mathematics and Theoretical Physics,
University of Cambridge, Wilberforce Road, Cambridge CB3 OWA, UK}
\author{Blake Sherwin}
\affiliation{Department of Applied Mathematics and Theoretical Physics,
University of Cambridge, Wilberforce Road, Cambridge CB3 OWA, UK}
\affiliation{Kavli Institute for Cosmology Cambridge, Madingley Road, Cambridge CB3 0HA, UK}

\begin{abstract}

Extragalactic foregrounds in Cosmic Microwave Background (CMB) temperature maps lead to significant biases in CMB lensing reconstruction if not properly accounted for.
Combinations of multifrequency data have been used to minimize the overall map variance (internal linear combination, or ILC), or specifically null a given foreground, but these are not tailored to CMB lensing.
In this paper, we derive an optimal multifrequency combination to jointly minimize CMB lensing noise and bias.
We focus on the standard lensing quadratic estimator, as well as the ``shear-only'' and source-hardened estimators, whose responses to foregrounds differ.
We show that an optimal multifrequency combination is a compromise between the ILC and joint deprojection, which nulls the thermal Sunyaev-Zel'dovich (tSZ) and Cosmic Infrared Background (CIB) contributions. 
In particular, for a Simons Observatory-like experiment with $\ell_{\text{max},T}=3000$, we find that profile hardening alone (with the standard ILC) reduces the bias to the lensing power amplitude by $40\%$, at a $20\%$ cost in noise, while the bias to the cross-correlation with a LSST-like sample is reduced by nearly an order of magnitude at a $10\%$ noise cost, relative to the standard quadratic estimator. 
With a small amount of joint deprojection the bias to the profile hardened estimator can be further reduced to less than half the statistical uncertainty on the respective amplitudes, at a $20\%$ and $5\%$ noise cost for the auto- and cross-correlation respectively, relative to the profile hardened estimator with the standard ILC weights. 
Finally, we explore possible improvements with more aggressive masking and varying $\ell_{\text{max,}T}$.
\end{abstract}

\maketitle

\section{Introduction}
CMB lensing \cite{1987A&A...184....1B,2006PhR...429....1L, 2010GReGr..42.2197H} allows us to reconstruct the projected mass distribution all the way to the surface of last scattering, enabling cosmological inference (e.g. neutrino masses, dark matter and dark energy) using a very well-understood source redshift and redshift weighting.  Extragalactic foregrounds, which are correlated with the lensing potential, are likely to be one of the main limiting factors \cite{2014ApJ...786...13V,2014JCAP...03..024O, 2018PhRvD..98b3534M,2018PhRvD..97b3512F,2019PhRvL.122r1301S,2019PhRvD..99b3508B, sailer_paper1} in temperature-based CMB lensing reconstruction, producing significant biases to the standard minimum variance quadratic estimator (QE, \cite{2002ApJ...574..566H}), which in turn bias both the reconstructed lensing power spectrum and the cross-correlations with tracers of the large-scale structure. In this paper we consider two techniques to mitigate these biases: geometric-based methods that distinguish lensing from foregrounds through their different spatial symmetries (shear-only reconstruction \cite{2019PhRvL.122r1301S} and bias-hardening \cite{sailer_paper1, 2014JCAP...03..024O}), and multifrequency based methods that exploit the different frequency scaling of the CMB and extragalactic foregrounds \cite{2009LNP...665..159D,2018PhRvD..98b3534M}. In particular, we explore the \textit{lensing-optimized} linear combinations of frequency maps to minimize a combination of bias and noise, for the standard QE, as well as for shear and bias-hardened estimators. 
Another promising method to control foreground biases to CMB lensing is the gradient-cleaned \cite{2018PhRvD..98b3534M} or symmetrized estimator \cite{2020arXiv200401139D, Patil:2019zbm}. We explore these methods in a companion paper \cite{DarwishInPrep}.
The remainder of the paper is organized as follows: in \S\ref{sec:CMB lensing quadratic estimators} we introduce the standard, shear-only, and bias-hardened quadratic estimators; in \S\ref{sec:review ilc} we review multifrequency methods for noise and bias mitigation; in \S\ref{sec:reconstruction_noise} (\ref{sec:reconstruction_bias}) we outline the calculation of the noise (bias) of a general quadratic estimator; in \S\ref{sec:results} we combine geometric- and frequency-based techniques to minimize a combination of lensing noise and bias; we conclude in \S\ref{sec:conclusions}. 
 
\section{CMB lensing quadratic estimators}
\label{sec:CMB lensing quadratic estimators}

In the presence of a foreground $s_{\bm{\ell}}$, the observed temperature covariance receives off-diagonal contributions from lensing and the foreground\footnote{$T_{\bm{\ell}} = T^\text{CMB}_{\bm{\ell}} + s_{\bm{\ell}}$, where $T^\text{CMB}_{\bm{\ell}}$ is the \textit{lensed} primary CMB, and the average is taken over realizations of the unlensed CMB at fixed $\kappa_{\bm{L}}$ and $s_{\bm{L}}$.} (see Eqs.~3-7 in \cite{sailer_paper1}):
\beq
\langle T_{\boldsymbol{\ell}} T_{\boldsymbol{L}-\boldsymbol{\ell}} \rangle
=
f^\kappa_{\boldsymbol{\ell},\boldsymbol{L}-\boldsymbol{\ell}}\kappa_{\boldsymbol{L}}
+
f^s_{\boldsymbol{\ell},\boldsymbol{L}-\boldsymbol{\ell}}s_{\boldsymbol{L}}
\label{eq:off_diagonal_covariance}
\eeq
to lowest order in $\kappa$ and $s$. 
The standard quadratic estimator (QE, \cite{2002ApJ...574..566H}) neglects the foreground term and seeks the minimum variance solution for $\kappa_\vL$ in Eq.~\eqref{eq:off_diagonal_covariance}. By neglecting the foreground contribution, it can lead to a biased answer \cite{2019PhRvL.122r1301S,sailer_paper1}.

On small scales where the foreground contributions are large, they are well approximated by a collection of unclustered halos or galaxies with azimuthally-symmetric profiles. 
In this regime, the shear-only $\kappa$ estimator, which is insensitive to such isotropic terms has been shown to significantly reduce all foreground-induced biases \cite{2019PhRvL.122r1301S}.

Bias-hardening \cite{2014JCAP...03..024O, 2013MNRAS.431..609N, sailer_paper1} seeks to measure the foreground contribution directly and ``subtract off'' the bias to the lensing convergence $\kappa_\vL$.
Since both $\kappa_{\boldsymbol{L}}$ and $s_{\boldsymbol{L}}$ produce off-diagonal covariances, the standard quadratic estimators for $\kappa$ and $s$ are biased \cite{2014JCAP...03..024O, sailer_paper1}:
\beq
    \begin{pmatrix}
     \langle\hat{\kappa}_{\boldsymbol{L}}\rangle\\
     \langle\hat{s}_{\boldsymbol{L}}\rangle
    \end{pmatrix} =
    \begin{pmatrix}
    1 & N^\kappa_{\boldsymbol{L}}\mathcal{R}_{\boldsymbol{L}}\\
    N^{s}_{\boldsymbol{L}}\mathcal{R}_{\boldsymbol{L}} & 1
    \end{pmatrix}
    \begin{pmatrix}
     \kappa_{\boldsymbol{L}}\\
     s_{\boldsymbol{L}}
    \end{pmatrix}
\label{eq:responses_k_s}
\eeq
where the response $\mathcal{R}_\vL$ is defined in \cite{sailer_paper1}. 
A simple matrix inversion leads to unbiased estimates of the lensing and foreground:
\begin{equation}
    \begin{pmatrix}
     \hat{\kappa}^\text{BH}_{\boldsymbol{L}}\\
     \hat{s}^{\text{BH}}_{\boldsymbol{L}}
    \end{pmatrix} =
    \begin{pmatrix}
    1 & N^\kappa_{\boldsymbol{L}}\mathcal{R}_{\boldsymbol{L}}\\
    N^{s}_{\boldsymbol{L}}\mathcal{R}_{\boldsymbol{L}} & 1
    \end{pmatrix}^{-1}
    \begin{pmatrix}
     \hat{\kappa}_{\boldsymbol{L}}\\
     \hat{s}_{\boldsymbol{L}}
    \end{pmatrix}
    .
\label{eq:bias-hardened_k_s}
\end{equation}
We assume that the foreground $s$ can be approximated by a collection of sources with identical profiles $u_{\bm{\ell}}$ \cite{sailer_paper1}. Explicitly, we assume $s_{\bm{\ell}} =\sum_i s_i e^{i \bm{\ell}\cdot\bm{x}_i}u_{\bm{\ell}}$, where $s_i$ $(\bm{x}_i)$ is the amplitude (position) of the $i$'th source. In this work we consider either hardening against point sources \cite{2014JCAP...03..024O, 2013MNRAS.431..609N, sailer_paper1} (point source hardening, or PSH), where the profile is taken to be a delta function ($u_{\bm{\ell}} = 1$), or against a tSZ-like profile \cite{sailer_paper1} (profile hardening, or PH), where the profile is taken to be the square root of the tSZ power spectrum at 150 GHz.

\section{Internal Linear Combination (ILC) Methods}
\label{sec:review ilc}

Let $(\boldsymbol{T}_{\bm{\ell}})_i$ denote the observed map in the $i$-th frequency channel. We assume that $\boldsymbol{T}_{\bm{\ell}}$ can be written in the form:
\begin{equation}
    \boldsymbol{T}_{\bm{\ell}}
    =
    \boldsymbol{1}
    T^\text{CMB}_{\bm{\ell}}
    +
    \boldsymbol{n}_{\bm{\ell}}
    +
    \sum_s \bm{A}_s s_{\bm{\ell}},
\label{eq:T_decomposition}
\end{equation}
where $\boldsymbol{1}=(1,1,\cdots,1)^T$ and $T^\text{CMB}_{\bm{\ell}}$ is the \textit{lensed} CMB. The ``noise'' $\bm{n}_{\bm{\ell}}$ receives contributions from the detectors' thermal fluctuations, the atmosphere, and galactic dust. The extragalactic foregrounds $s_{\bm{\ell}}$ include tSZ, kSZ, CIB, and radio point sources. In Eq.~\eqref{eq:T_decomposition} we assume that each map has been calibrated to have unit response to the lensed CMB. 
The lensed CMB in each frequency channel is thus determined by a single template $T^\text{CMB}_{\bm{\ell}}$. 
Similarly, we assume that each foreground $s$ follows a single spatial template $s_{\bm{\ell}}$, up to a frequency dependence $\bm{A}_s$.
Our normalized linear estimator for the lensed CMB takes the form $\hat{T}_{\bm{\ell}} = \bm{w}_\ell^T \bm{T}_{\bm{\ell}}$ for some weights $\bm{w}_\ell$, which we are free to choose subject to the constraint $\bm{w}_\ell^T\bm{1} = 1$. The noise of the estimator $\hat{T}_\ell$ is 
\begin{equation}
    C^\text{tot}_{\ell} = \bm{w}^T_\ell \bm{C}_\ell \bm{w}_\ell,
\end{equation}
where $(2\pi)^2 \delta^D_{\bm{0}}  \bm{C}_\ell = \langle
\bm{T}_{\bm{\ell}}
\bm{T}_{-\bm{\ell}}^T
\rangle$
is the covariance matrix of the maps $\bm{T}_{\bm{\ell}}$.

In this work we consider a Simons Observatory-like experiment, which will observe the CMB in six frequency channels: $27$, $39$, $93$, $145$, $225$, and $280$ GHz. We follow \cite{2019JCAP...02..056A} in modeling the detector noise and atmospheric contributions to $\bm{C}_{\bm{\ell}}$, assuming the ``goal'' noise levels. The theory power spectra and SEDs for the galactic dust and extragalactic foregrounds are taken from \cite{2013JCAP...07..025D}.

Before deriving the optimal linear combination for CMB lensing, we start by reviewing the case of the standard ILC and deprojection.
These two limiting cases, commonly used in CMB data analysis, correspond to two extremes. 
The standard ILC focuses only on the total map noise, without trying to reduce any specific foreground bias,
whereas deprojection completely nulls a given foreground (in so far as its frequency dependence is known exactly), regardless of the noise penalty.
As we show below, our CMB lensing-optimized ILC is a compromise between these two extremes. 

\subsection{Standard ILC}

The standard harmonic ILC is defined to minimize the total map variance from noise and foregrounds. 
Minimizing $C^\text{tot}_\ell$ subject to the constraint $\bm{w}^T_\ell\bm{1}=1$ can be achieved by introducing a Lagrange multiplier $\lambda_\ell$ and minimizing the Lagrangian:
\beq
\mathcal{L}\left[ w_\ell, \lambda_\ell \right]
\equiv
C^\text{tot}_\ell
+
\lambda_\ell
\left( 1 - \bm{w}^T_\ell\bm{1} \right).
\eeq
Setting $\partial \mathcal{L}/\partial \bm{w}_\ell=0$ forces $\bm{w}_\ell \propto \bm{C}_\ell^{-1}\bm{1}$. The normalization is fixed by the constraint $\bm{w}_\ell^T \bm{1} = 1$, from which we recover the standard result \cite{Eriksen_2004,Tegmark_2003}:
\beq
\bm{w}^\text{ILC}_{\ell}
=
\frac{
\bm{C}^{-1}_\ell  \bm{1}}
{\bm{1}^T\bm{C}^{-1}_\ell  \bm{1}}.
\label{eq: ILC_weights}
\eeq
While the ILC minimizes the total power spectrum of the combined map, there is no guarantee that the power spectrum of individual foreround components will be reduced, so long that the foregrounds are a subdominant source of power. Even if the foreground power spectra are reduced, the ILC it is not designed to reduce the bispectrum or trispectrum of the foregrounds, which in turn source the biases to lensing. Thus it is possible for a foreground to have a subdominant power spectrum, but a dominant bispectrum or trispectrum, in which case the ILC does nothing to suppress the resulting bias.

\subsection{Deprojection}

\begin{figure}
    \centering
    \includegraphics[width=\linewidth]{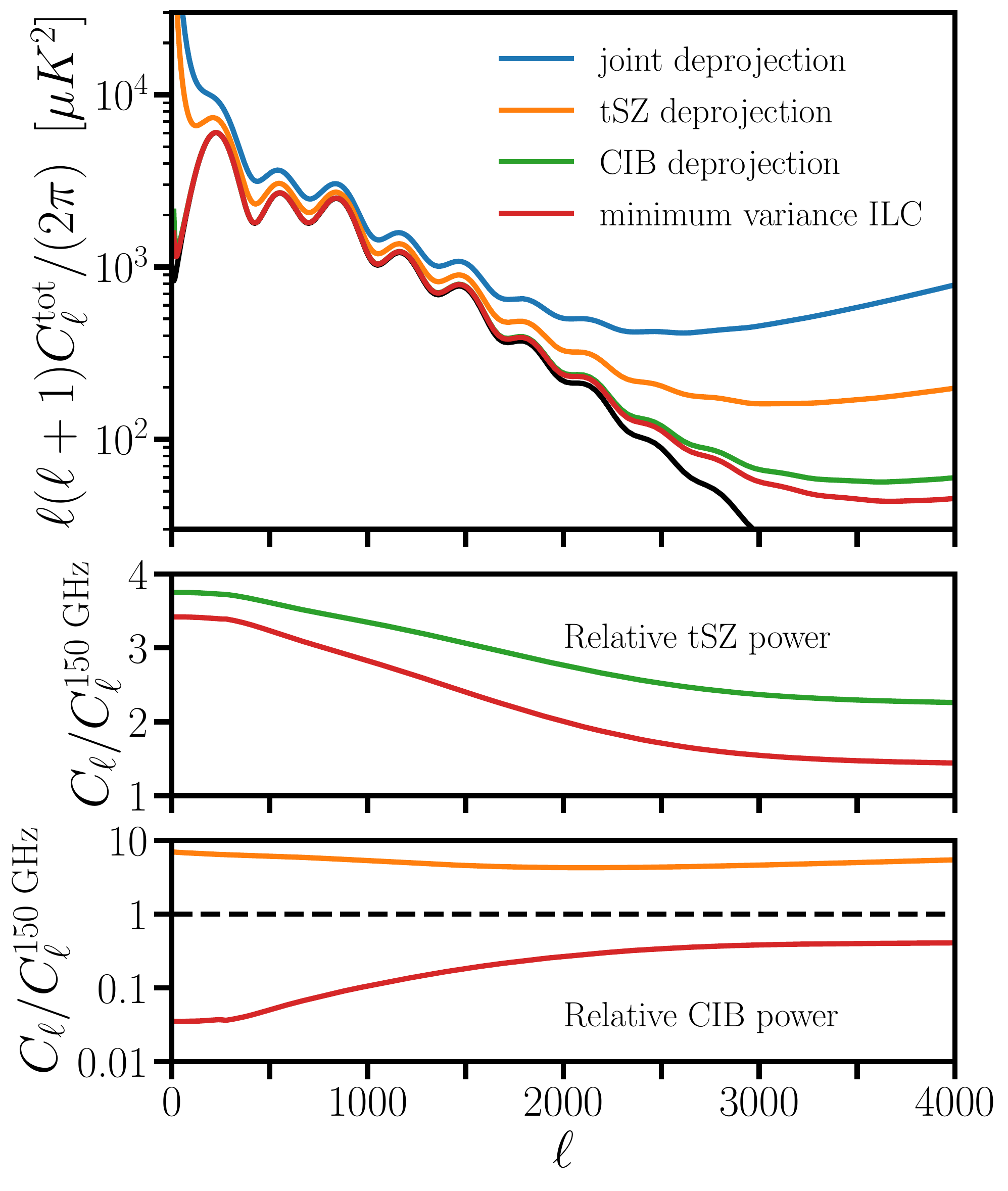}
    \caption{
    The top panel shows the temperature power spectrum for different multifrequency combinations, compared to the noiseless lensed CMB (black curve).
    Compared to the minimum-variance combination (ILC, red), the deprojection of tSZ (orange) comes at a larger cost in noise than that of CIB (green).
    Jointly deprojecting tSZ and CIB (blue) causes a more than ten-fold noise increase at $\ell\sim 3000$, where most of the lensing information resides.
    We note that the significant increase in power at large scales from joint or tSZ deprojection is primarily due to a large increase in atmospheric noise.
    The middle and bottom panels show the tSZ and CIB power spectra, respectively, compared to the single-frequency case (150~GHz).
    For tSZ (middle), the standard ILC and the CIB deprojection cause a significant increase.
    For CIB (bottom), the ILC very effectively reduces the power; however, the tSZ-deprojection enhances it by more than an order of magnitude.
    }
    \label{fig:various_ILC_noises}
\end{figure}

In certain applications, one may need to null a specific foreground. 
For instance, tSZ measurements ($y$-maps and stacks) are contaminated by thermal dust emission, if unaccounted for. In this context, one may use constrained ILC maps with CIB deprojection \cite{Remazeilles_2010,Madhavacheril_2020}.
As shown in Fig.~\ref{fig:various_ILC_noises},
one can deproject one or several foregrounds to remove the corresponding biases, at the cost of increasing the temperature map noise. 
Deprojection amounts to choosing the weights $\bm{w}_{\ell}$ to have zero response to foreground $s$: $\bm{w}^T_{\ell} \bm{A}_s=0$. Minimizing the map noise subject to this additional constraint leads to minimizing  
\beq
\bal
\mathcal{L}\left[ w_\ell, \lambda_\ell, \sigma_\ell \right]
\equiv\,
&C^\text{tot}_\ell
+
\lambda_\ell
\left( 1 - \bm{w}^T_\ell\bm{1} \right)- \sigma_\ell \bm{w}_\ell^T \bm{A}_s
.
\eal
\eeq
Setting 
$\partial{\mathcal{L}} / \partial{\bm{w}_\ell} = 0$
forces $\bm{w}_\ell \propto \bm{C}^{-1}_\ell\left( \bm{1} +\sigma_\ell \bm{A}_s/\lambda_\ell\right)$. Enforcing the constraint $\bm{w}^T_{\bm{\ell}} \bm{A}_s=0$ fixes the coefficient:
\beq
\sigma_\ell/\lambda_\ell 
=
-
\frac{
\bm{1}^T\bm{C}_\ell^{-1}\bm{A}_s
}{
\bm{A}^T_s \bm{C}_\ell^{-1} \bm{A}_s
}.
\eeq 
Finally, $\bm{w}_\ell^T\bm{1}=1$ fixes the normalization.
This technique can be trivially extended to simultaneously deproject more than one foreground ($\bm{w}^T_{\ell} \bm{A}_{s_i}=0$ for $i=1,\cdots,n$) by introducing additional Lagrange multipliers.

\section{Reconstruction Noise}
\label{sec:reconstruction_noise}

A general quadratic estimator $\hat{\kappa}_{\bm{L}}$ of the 
lensing convergence $\kappa_{\bm{L}}$ takes the form
\begin{equation}
    \hat{\kappa}_{\bm{L}}[T,T'] 
    = 
    N_{\boldsymbol{L}} 
    \int \frac{d^2\boldsymbol{\ell}}{(2\pi)^2}
    F_{\boldsymbol{\ell},\boldsymbol{L}-\boldsymbol{\ell}}
    T_{\boldsymbol{\ell}}
    T'_{\boldsymbol{L}-\boldsymbol{\ell}}
\label{eq:estimator}
\end{equation}
where $N_{\boldsymbol{L}}$ is some normalization, chosen so that the estimator has unit response to the signal $\langle \hat{\kappa}_{\bm{L}}\rangle = \kappa_{\bm{L}} + \text{ bias}$, and $F_{\boldsymbol{\ell},\boldsymbol{L}-\boldsymbol{\ell}}$ are the weights, which are in principle arbitrary. 
In this work, both maps $T$ and $T'$ have unit response to the CMB, such that the normalization is
\begin{equation}
    N_{\bm{L}}^{-1} = \int \frac{d^2\bm{\ell}}{(2\pi)^2} F_{\bm{\ell},\bm{L}-\bm{\ell}} f^\kappa_{\bm{\ell},\bm{L}-\bm{\ell}}.
\end{equation}
This estimator applied to the map $\hat{T}_{\bm{\ell}}$ has noise

\begin{equation}
\begin{aligned}
    \mathcal{N}_L[\bm{w}]
    &=
    \langle \hat{\kappa}_{\bm{L}}[\hat{T},\hat{T}]\hat{\kappa}_{-\bm{L}}[\hat{T},\hat{T}]\rangle\big/(2\pi)^2 \delta^D_{\bm{0}}
    \\
    &=
    2
    N^2_{\boldsymbol{L}} 
    \int \frac{d^2\boldsymbol{\ell}}{(2\pi)^2}
    F^2_{\boldsymbol{\ell},\boldsymbol{L}-\boldsymbol{\ell}}
    C^\text{tot}_\ell
    C^\text{tot}_{|\bm{L}-\bm{\ell}|}
\label{eq:general_noise}
\end{aligned}
\end{equation}
for $L>0$. 
In going from the first to second line of Eq.~\eqref{eq:general_noise}, we have only kept the disconnected contributions to the auto-correlation, as is customary in lensing noise $N^{(0)}$ calculations.
We also assumed that $F$ is symmetric and even in both of its arguments.
\begin{figure}[h!]
    \centering
    \includegraphics[width=\linewidth]{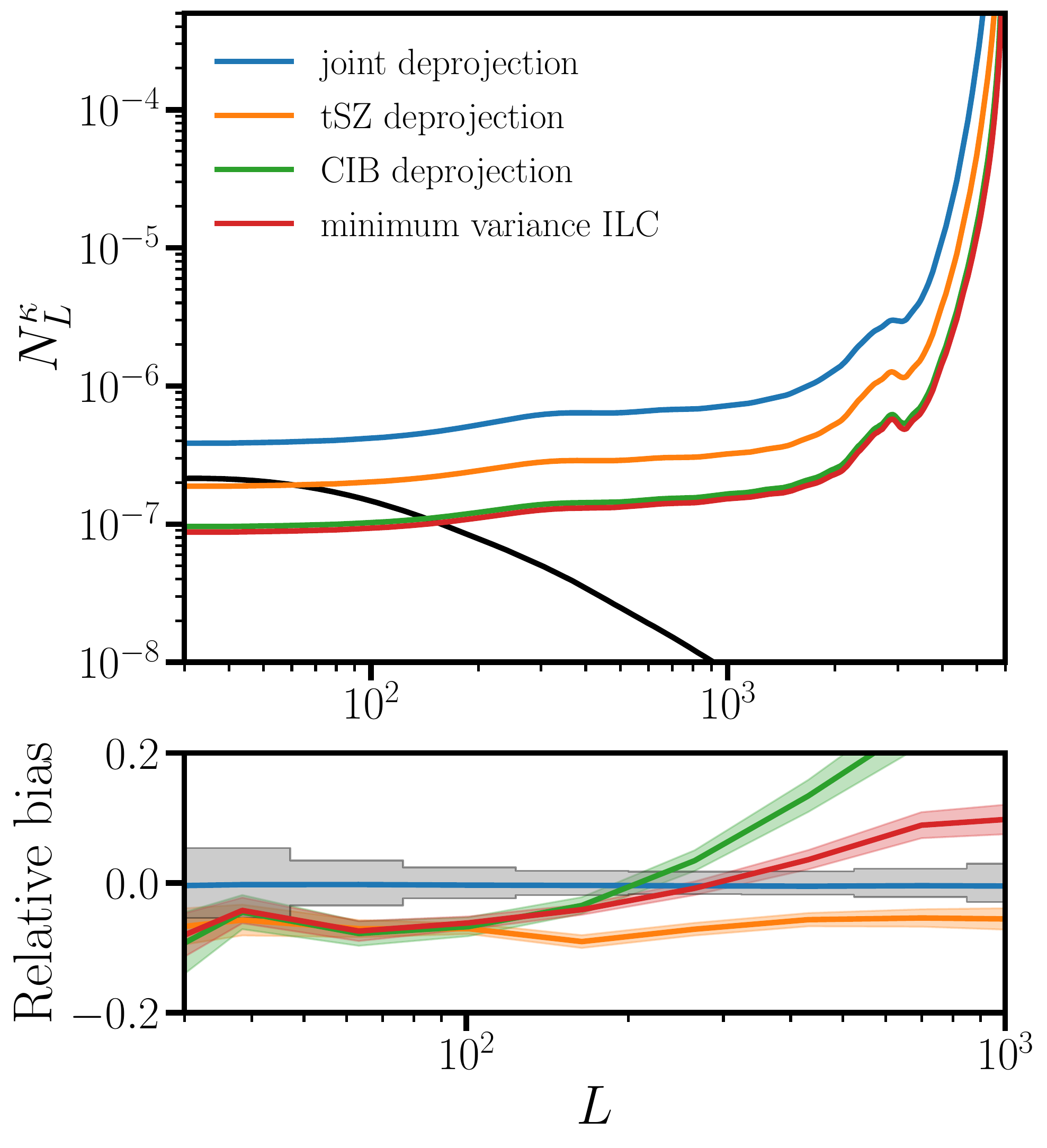}
    \caption{
    As illustrated in the top panel, the standard ILC is the linear combination of multifrequency maps which minimizes the lensing noise, for the standard QE and all the other estimators considered in this paper.
    As expected from Fig.~\ref{fig:various_ILC_noises}, deprojecting CIB (green) comes at almost no cost in lensing noise, whereas deprojecting tSZ (yellow) multiplies the noise by $\sim 2$.
    Despite causing a more than ten-fold increase in temperature noise, joint deprojection only causes a four-fold increase in lensing noise. 
    This can be understood by counting the number of signal-dominated temperature modes.
    The bottom panel shows the fractional bias to the CMB lensing auto-spectrum for the various frequency combinations.
    The gray boxes denote the bandpower errors for the standard ILC. 
    For the ILC, the bias is largely dominated by the enhanced tSZ signal. 
    While the joint deprojection successfully nulls all biases, simply deprojecting tSZ or CIB does not.
    }
    \label{fig:fiducial_lensing_noise_bias}
\end{figure}

Note that both the normalization $N_{\bm{L}}$ and the weights $F_{\bm{\ell},\bm{L}-\bm{\ell}}$ could in principle depend on the weights $\bm{w}_{\bm{\ell}}$.
In this analysis, the spatial weights $F_{\bm{\ell},\bm{L}-\bm{\ell}}$ depend on the map noise $C^\text{tot}_\ell$, which in turn depend on the frequency weights $\bm{w}_\ell$.
In App.~\ref{app:ilc_minimizes_lensing_noise}, we show that the ILC is the frequency combination which minimizes the lensing noise, for all the estimators we consider here.
The top panel of Fig.~\ref{fig:fiducial_lensing_noise_bias} shows the lensing noise for the standard QE for various frequency combinations, with $\ell_{\text{max,}T}=3000$.

\subsection{Lensing amplitude noise}

The lensing amplitude $A_\text{lens}$ is defined to be the ratio of the measured convergence auto-spectrum to the truth, with fiducial value $1$. An estimator for the lensing amplitude can be easily constructed for each mode $L$: $\hat{A}_{\text{lens},L}= \hat{C}^{\kappa}_L/C^\kappa_L$, where $\hat{C}^{\kappa}_L$ denotes the measured convergence auto-spectrum. Given a range of modes, the minimum variance estimator takes the form
\beq
\hat{A}_\text{lens} = 
\left(
\int \frac{d^2 \bm{L}}{(2\pi)^2}
\frac{\hat{C}^{\kappa}_L}{C^{\kappa}_L} 
\frac{(C^\kappa_L)^2}{\sigma^2_L}
\right)
\bigg/
\left(
\int \frac{d^2 \bm{L}}{(2\pi)^2}
\frac{(C^\kappa_L)^2}{\sigma^2_L}
\right),
\eeq
where
$\sigma^2_L = 2 ( C_L^\kappa + N_L^\kappa )^2$. This is simply an inverse variance weighting of the estimators for each Fourier mode.

Here we simply aim to minimize the lensing noise:
\beq
\sigma_{A_\text{lens}}^2
=
\frac{1}{4\pi f_\text{sky}}
\left(
\int \frac{d^2 \bm{L}}{(2\pi)^2}
\frac{(C_L^{\kappa})^2}{ \sigma^2_L}
\right)^{-1}
\label{eq:amplitude_noise}
\eeq
Minimizing $\sigma_{A_\text{lens}}$ is equivalent to maximizing the integral in Eq.~\eqref{eq:amplitude_noise}, which has a strictly positive integrand. Thus maximizing the integral is equivalent to maximizing the integrand for each $\bm{L}$, which in turn is equivalent to minimizing $\sigma_L^2$, or $N^\kappa_L$. Therefore the weights that minimize the noise of the lensing amplitude $\sigma_{A_\text{lens}}$ are equivalent to the weights that minimize the lensing reconstruction noise $N^\kappa_L$, which is given by the standard ILC.

\begin{figure*}
    \centering
    \includegraphics[width=0.48\linewidth]{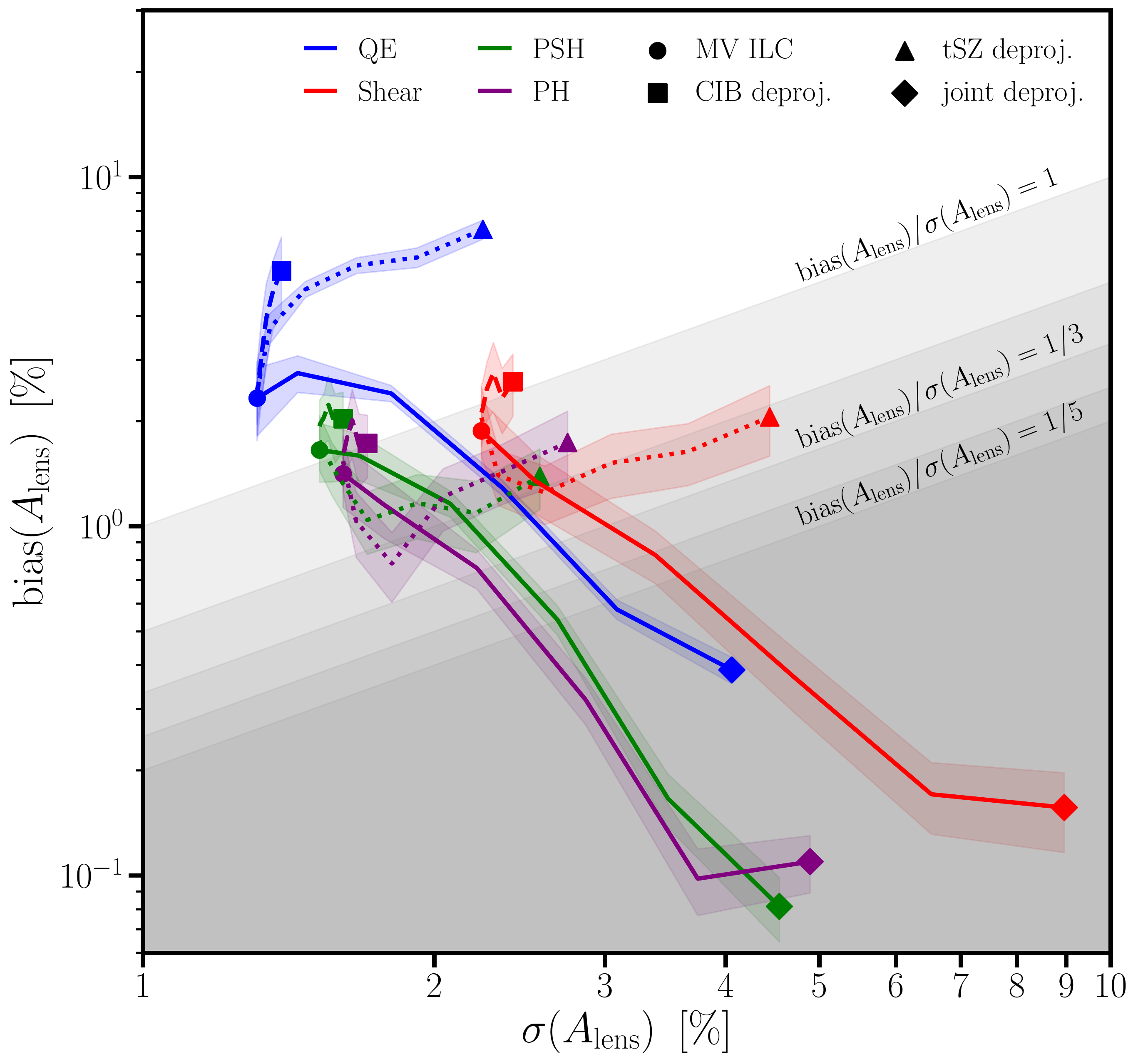}
    \includegraphics[width=0.48\linewidth]{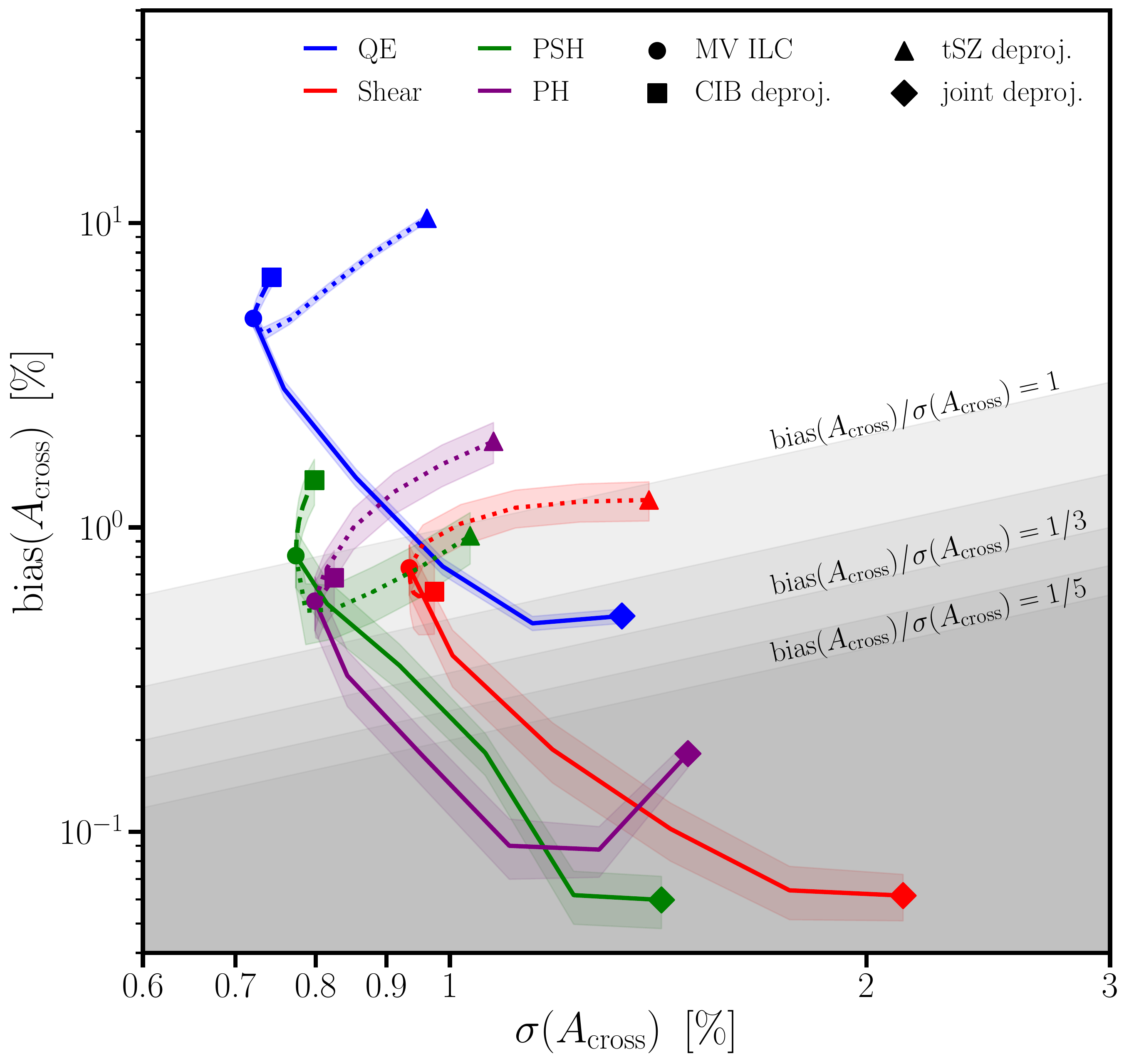}
    \caption{
    \textit{Left:} The total $|$bias$|$ and noise on the lensing amplitude $A_\text{lens}$ for the various ILC combinations. The biases for the standard QE are shown in blue, Shear in red, the Point Source Hardened estimator in green, and the Profile Hardened estimator in purple. \textit{Right:} The same as the left panel, but for the amplitude $A_\text{cross}$ of the cross-correlation with the mock LSST sample. Circles denote the minimum variance (MV) ILC, squares (triangles) denote CIB (tSZ) deprojection, while diamonds denote the simultaneous deprojection of both the CIB and tSZ. The lines connecting MV ILC to either CIB, tSZ or joint deprojection follow Eq.~\eqref{eq:walks}, with step sizes $\Delta t = 0.2$. Throughout we sum over $20<L<1000$ when calculating the bias and noise for either $A_\text{lens}$ or $A_\text{cross}$, and take $f_\text{sky}=0.4$.}
    \label{fig:money_plot}
\end{figure*}

\section{Bias to lensing reconstruction from simulations}
\label{sec:reconstruction_bias}

\subsection{Method}

To estimate the non-Gaussian lensing biases from extragalactic foregrounds, we follow the procedure in \cite{sailer_paper1, 2019PhRvL.122r1301S}.
We ignore the lensed-foreground bias, due to the fact that the foregrounds are emitted at cosmological distances and are themselves lensed \cite{2019PhRvD.100l3504M}.
The non-Gaussian foreground bias to the lensing auto-spectrum can be decomposed into a primary, secondary and trispectrum terms \cite{2019PhRvL.122r1301S, 2014JCAP...03..024O, 2018PhRvD..97b3512F, 2014ApJ...786...13V}:
\beq
\bal
\langle \mathcal{Q} \mathcal{Q} \rangle
&=
\underbrace{\langle \kappa_\text{CMB} \kappa_\text{CMB} \rangle}
_\text{lensing signal}
+
\underbrace{2\langle \mathcal{Q}[T^\text{CMB}, T^\text{CMB}] \mathcal{Q}[s, s] \rangle}
_\text{Primary bispectrum bias}\\
&+
\underbrace{4\langle \mathcal{Q}[T^\text{CMB}, s] \mathcal{Q}[T^\text{CMB}, s] \rangle}
_\text{Secondary bispectrum bias}
+
\underbrace{\langle \mathcal{Q}[s, s] \mathcal{Q}[s, s] \rangle}
_\text{Trispectrum bias},\\
\eal
\label{eq:auto_biases}
\eeq
where $\mathcal{Q}[T,T']$ represents any (symmetric) quadratic estimator evaluated on the maps $T$ and $T'$. For the cross-correlation of CMB lensing with an external tracer $g$ of the mass (e.g., galaxies or galaxy lensing), only the primary bispectrum bias appears:
\beq
\langle g \mathcal{Q} \rangle
= 
\underbrace{\langle g \kappa_\text{CMB} \rangle}
_\text{lensing signal}
+
\underbrace{\langle g \mathcal{Q}[s, s] \rangle}
_\text{bispectrum bias}.
\eeq
In both cases, the bispectrum biases (primary and secondary) appear because the foregrounds are simultaneously non-Gaussian and correlated with the true CMB lensing signal.
The trispectrum bias in auto-correlation is present for any non-Gaussian foreground, regardless of any potential correlation with CMB lensing.

In what follows, we estimate each of these bias terms separately for the total contribution from extragalactic foregrounds (tSZ $+$ CIB $+$ kSZ $+$ radio PS).
We apply the method presented in \cite{2019PhRvL.122r1301S} identically.
This uses the non-Gaussian foreground simulations of \cite{2010ApJ...709..920S} at 150~GHz, which are then rescaled to any other frequency, given the expected SED of each component from \cite{2013JCAP...07..025D}, as implemented in \texttt{MicroCoSM}\footnote{\url{https://github.com/EmmanuelSchaan/MicroCoSM}}. We choose the tracer $g$ to be a LSST-like sample constructed from reweighting the simulated halos, as in \cite{2019PhRvL.122r1301S,sailer_paper1}, to obtain the correct redshift distribution and linear bias.

In all cases, the foreground maps from \cite{2010ApJ...709..920S} are masked before the lensing reconstruction.
The mask is obtained by detecting the individual point sources with flux density higher than 5~mJy in the 150~GHz map using a matched filter, and masking the foreground map with a 3 arcmin radius disks around each source. 
We then add the unmasked lensed CMB map to the foreground map.
This simulates the effect of inpainting, or source template subtraction, as expected for SO.
We also explore a more aggressive masking strategy in App.~\ref{app:mask}, which could reduce the bias to the profile hardened estimator by as much as $30\%$, provided that the effects of the mean field produced by this more aggressive mask are kept under control. Masking the point sources may also mask the high $\kappa$ peaks, which would bias the signal low \cite{Fabbian_2021,Liu_2015}, but inpainting or template subtraction should circumvent this problem.

Another important method for avoiding foreground contamination is to discard the highest multipoles in the input temperature map.
Unless otherwise indicated, we assume that only modes with $\ell \leq \ell_{\text{max,}T}=3000$ are included.
In App.~\ref{app:lmaxt}, we explore the sensitivity of our results to this choice.

\subsection{Lensing amplitude bias}

The bias to the lensing amplitude $A_\text{lens}$ is given by
\beq
\text{bias}\left( A_\text{lens} \right)
=
\frac{
\int d^2\bm{L}\,
C_L^\kappa\
\text{bias}\left( C_L^\kappa \right)
/\sigma^2_L
}
{
\int d^2 \bm{L}\,
(C_L^\kappa)^2
/\sigma^2_L
}
,
\label{eq:amplitude_bias}
\eeq
where the bias to the lensing auto-correlation is the sum of the primary, secondary and trispectrum terms, as defined in Eq.~\eqref{eq:auto_biases}. Analogous expressions for the noise and bias of the amplitude $A_\text{cross}$ of the cross-correlation of CMB lensing with an external tracer $g$ may be obtained by substituting $C^\kappa_L$ with $C^{\kappa g}_L $ and setting $\sigma^2_L = (C^\kappa_L+N^\kappa_\ell)(C^g_L+N^g_L)+(C^{\kappa g}_L)^2$ in Eqs. \eqref{eq:amplitude_noise} and \eqref{eq:amplitude_bias}.

We note that cancellations could potentially occur in Eq.~\eqref{eq:amplitude_bias} between the biases at different lensing multipoles. In App.~\ref{app:app1} we explore how our optimal ILC weights change if we replace $\text{bias}(C^\kappa_L)$ with $|\text{bias}(C^\kappa_L)|$ in the expression above, which removes these cancellations.

\section{Minimizing lensing bias and noise}
\label{sec:results}

In this section we piece together the preceding results and show that an optimal multifrequency linear combination for minimizing a combination of noise and bias (for both $A_\text{lens}$ and $A_\text{cross}$) lies between the standard ILC and joint deprojection.

\subsection{Method}

Ideally, one would solve for the scale and frequency-dependent weights $\bm{w}_\ell$ which minimize some function $\mathcal{L}$ of the variance and bias of $A_\text{lens}$ (or $A_\text{cross}$). A reasonable choice for $\mathcal{L}$ is:
\beq
\mathcal{L}[\bm{w}_\ell] = \sigma^2(A_\text{lens}) + f_b^2 \text{bias}^2(A_\text{lens}),
\label{eq:ideal_loss}
\eeq
where $f_b$ is a tuning parameter for favoring either noise $(f_b=0)$ or bias $(f_b\to\infty)$.
In practice, however, solving for $\bm{w}_\ell$ via a brute force minimization of $\mathcal{L}$ is computationally intractable.
The problem can be greatly simplified with a few well-motivated approximations; however, these turn out not to be accurate enough for our purposes (see App.~\ref{app:failed_approximations}).
We circumvent this issue with the following ansantz:
\beq
\bm{w}_\ell(t) =
      t\bm{X}_\ell + (1-t)\bm{w}^\text{ILC}_\ell,\\
\label{eq:walks}
\eeq
which smoothly connects the standard ILC to some form of deprojection ($\bm{X}_\ell = \bm{w}^\text{CIB deproj.}_\ell$, $\bm{w}^\text{tSZ deproj.}_\ell$, or $\bm{w}^\text{joint deproj.}_\ell$), and reduces the dimensionality of the parameter space from $5\times(\text{\# of }\ell\text{-bins})$ to $1$. In what follows we compute the bias and noise for each estimator along the line segment $t\in[0,1]$, in steps of $\Delta t = 0.2$. The optimal value of $t$ could in principle be found through minimizing Eq.~\eqref{eq:ideal_loss} for fixed $f_b$. However, the value of $f_b$ is arbitrarily chosen to reduce the bias to some desirable level, which can just as well be accomplished by choosing $t$ to be the smallest value such that $\text{bias}(A_\text{lens})/\sigma(A_\text{lens})$ is below some desirable cutoff. This is similar in spirit to the approach taken in deriving the partially constrained ILC \cite{Abylkairov_2021}. However, our cutoff is implemented directly at the bias level, while the cutoff in \cite{Abylkairov_2021} is implemented at the foreground power level.

\subsection{Results}

Our results are summarized in Fig.~\ref{fig:money_plot}, where we plot the bias vs noise for each estimator as we vary the weights $\bm{w}_\ell$ between the standard ILC and some form of deprojection $\bm{X}_\ell$, following the prescription of Eq.~\eqref{eq:walks}. We first note that deprojecting a single foreground (tSZ or CIB) often boosts the overall bias. As shown in bottom two panels Fig.~\ref{fig:various_ILC_noises}, deprojecting tSZ significantly boosts the CIB power spectrum (and vise versa), increasing the overall bias for both the auto and cross-correlation in many cases. 

We find that hardening against a tSZ-like profile outperforms the remaining estimators for any $\text{bias}/\text{noise}$ cutoff\footnote{This statement is no longer true near joint deprojection ($t\simeq 1$). In this regime, the PSH estimator has a lower bias (and noise) than the PH estimator, since in this case the foreground biases are primarily sourced by radio point sources (as opposed to tSZ or CIB), which the former estimator is designed to null. In practice, however, one does not need to push to $t\simeq 1$ to reduce the bias to an acceptable level.}, both in auto- and cross-correlation. Profile hardening at ILC ($t=0$, purple circles in Fig.~\ref{fig:money_plot}) reduces the bias to the lensing amplitude by $40\%$, at a $20\%$ cost in noise, while the bias to the cross-correlation is reduced by nearly an order of magnitude at a $10\%$ noise cost, relative to the standard quadratic estimator at ILC (blue circles in Fig.~\ref{fig:money_plot}). 

For all estimators, comfortably reducing the bias below $\sigma(A_\text{lens})/2$ (or $\sigma(A_\text{cross})/2$) requires partially deprojecting both the tSZ and CIB. We find that the noise cost for partial joint deprojection is much more significant for the auto-correlation than for the cross-correlation. In particular, for the profile hardened estimator, reducing the bias below $\sigma/2$ via partial joint deprojection corresponds to $t\sim 0.2$, resulting in a $20\%$ and $5\%$ noise penalty for the auto- and cross-correlation respectively, relative to profile hardening at ILC\footnote{Relative to the standard QE at ILC, the noise penalty for reducing the bias below $\sigma/2$ using profile hardening and partial joint deprojection is $50\%$ and $15\%$ for the auto- and cross-correlation respectively.}. Further improvements can be made by pushing to a higher $\ell_{\text{max},T}$ and moving closer to deprojection (higher $t$), or through a more aggressive masking scheme, as discussed in Apps.~\ref{app:lmaxt} and \ref{app:mask} respectively.

\section{Conclusions}
\label{sec:conclusions}
CMB lensing is quickly becoming one of the most powerful cosmological tools. Percent level measurements from the next generation of experiments require excellent control of extragalactic foregrounds, especially from temperature anisotropies.
Traditional analyses have used multifrequency foreground reduction methods when possible. Here we combined geometric methods with multifrequency information to optimize the lensing reconstruction. In doing so, we maximize the amount of information that can be extracted and the robustness of the measurement, while minimizing the potential biases. We also note that any multifrequency linear combination alone will not reduce the bias from kSZ, since it preserves the black-body nature of the CMB. While kSZ bias can be significant \cite{2018PhRvD..97b3512F}, it is quite effectively reduced by the shear-only or bias hardened estimators \cite{2019PhRvL.122r1301S, sailer_paper1}. Moreover, direct measurement of kSZ by combining low-redshift galaxy catalogs together with high-resolution CMB maps \cite{Schaan2020} can potentially provide templates for subtraction at the map level. We find that the standard ILC can lead to significant biases, which can be partially mitigated with geometric methods such as profile hardening, which alone can reduce the bias by $40\%$ (in auto-correlation) or almost an order of magnitude (in cross-correlation) with minimal ($10-20\%$) increase in noise. The bias can be further reduced to any desirable $\text{bias}/\text{noise}$ cutoff via partial joint deprojection of the tSZ and CIB, with the auto-correlation suffering a higher relative noise penalty (factor of $\sim4$) than the cross-correlation. We find that deprojection of tSZ or CIB alone do not reduce the biases, due to an enhancement in the component that is not deprojected. 
Finally, we have explored possible improvements with more aggressive masking (App.~\ref{app:mask}) and varying $\ell_{\text{max,}T}$ (App.~\ref{app:lmaxt}).

While here we have focused on temperature reconstruction, similar bias hardening techniques should be applicable to extragalactic foregrounds in polarization as well, where the biases are expected to be less severe. 
We leave a full exploration of this to future work. 

\section*{Acknowledgments}

We thank Anthony Challinor, Jo Dunkley, Giulio Fabbian, Colin Hill, Akito Kusaka, Antony Lewis, Mathew Madhavacheril, Neelima Sehgal, Alexander Van Engelen and Martin White for useful discussions in the preparation of this work. 
N.S. is supported by the NSF. 
E.S. is supported by the Chamberlain fellowship at Lawrence Berkeley National Laboratory. 
S.F. is supported by the Physics Division of Lawrence Berkeley National Laboratory. This work used resources of the National Energy Research Scientific Computing Center, a DOE Office of Science User Facility supported by the Office of Science of the U.S. Department of Energy under Contract No. DE-AC02-05CH11231.

\bibliographystyle{prsty.bst}
\bibliography{main}

\onecolumngrid
\appendix

\section{Cancellations}
\label{app:app1}

\begin{figure*}[!h]
    \centering
    \includegraphics[width=0.48\linewidth]{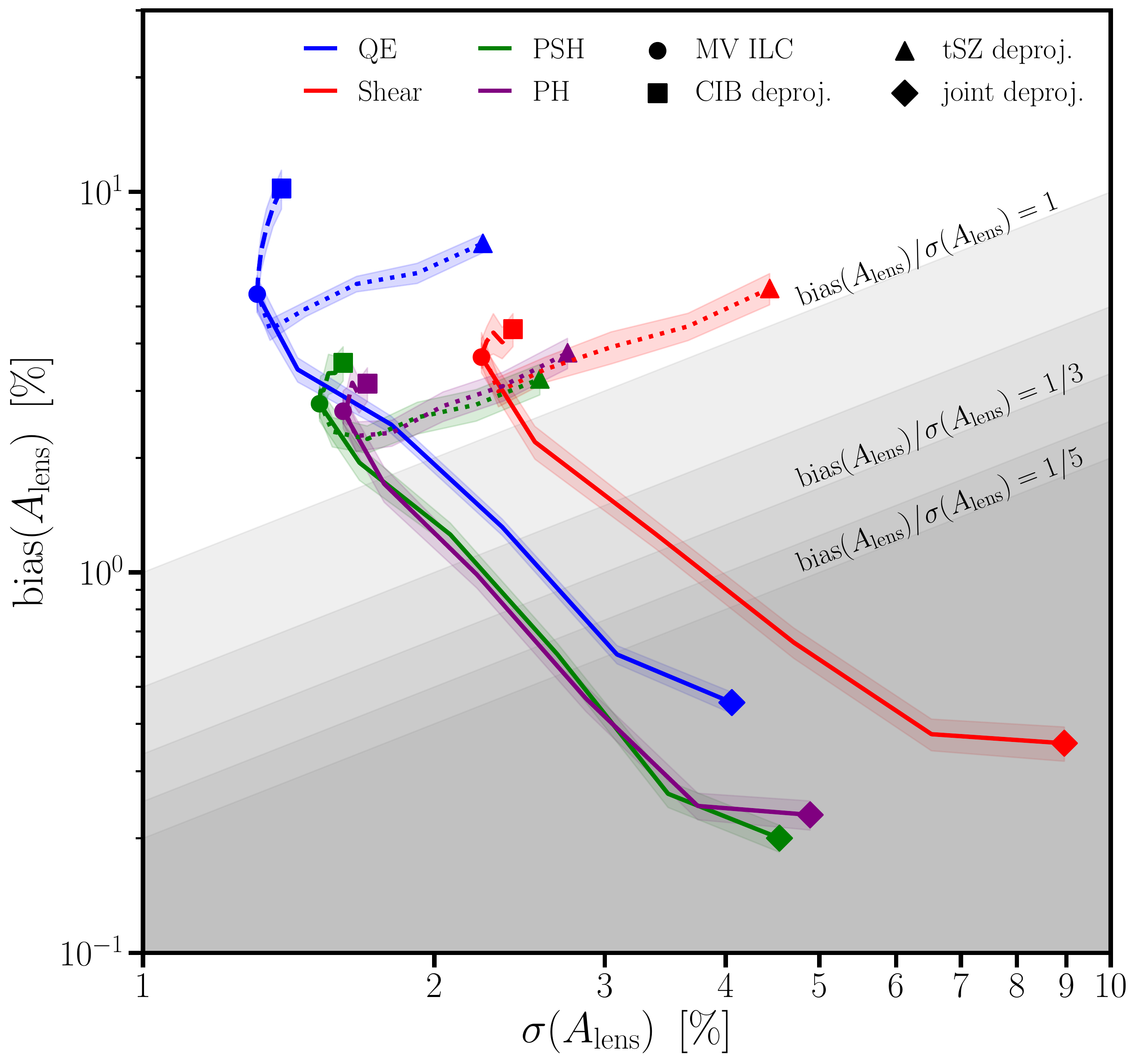}
    \includegraphics[width=0.48\linewidth]{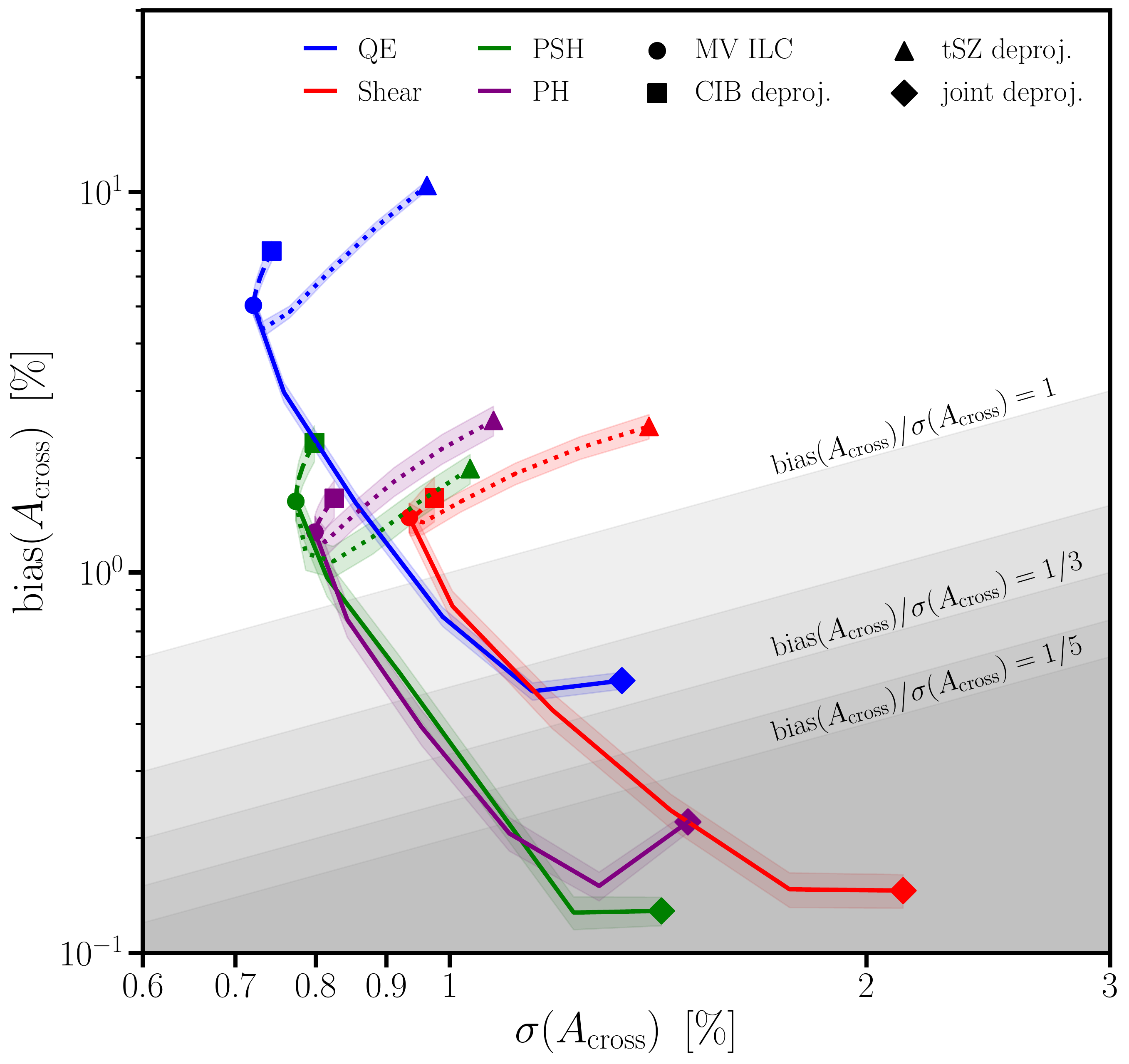}
    \caption{The same as Fig.~\ref{fig:money_plot}, but with $|\text{bias}(C^\kappa_L)|$ replacing $\text{bias}(C^\kappa_L)$ in Eq.~\eqref{eq:amplitude_bias}, and similarly for the cross-correlation with the mock LSST sample.}
    \label{fig:abs_bias}
\end{figure*}

As mentioned in \S\ref{sec:reconstruction_bias}, the definition of $\text{bias}(A_\text{lens})$ permits undesirable cancellations in biases across lensing multipoles. A simple method for avoiding these cancellations is to replace $\text{bias}(C^\kappa_L)$ with $|\text{bias}(C^\kappa_L)|$ in Eq.~\eqref{eq:amplitude_bias}. Fig.~\ref{fig:abs_bias} shows how our results change with this substitution. In particular for the profile hardened estimator, we find that one has to deproject out to $t\sim 0.4$ (as opposed to $t\sim 0.2$) in order to achieve bias/noise $<1/2$.

In addition to the aforementioned cancellations, for any given foreground, the primary, secondary and trispectrum terms may also cancel separately due to the linear combination of frequencies. While the distinction of primary, secondary and trispectrum terms is somewhat arbitrary, this cancellation relies on the relative sizes of the foreground bispectrum and trispectrum to be correct in the simulation used. Additionally, at any lensing multipole, the lensing biases from two distinct foregrounds may have different signs and partially cancel. While one can in principle take advantage of these cancellations, doing so relies on the simulations having the correct relative amplitude of each foreground, including their correlations. We will explore the sensitivity of our optimized ILC to these results in a future work, which will more realistically include polarization, in addition to temperature data.

\section{Proof that ILC minimizes the lensing noise for all QEs}
\label{app:ilc_minimizes_lensing_noise}

In this Appendix, we show that the linear combination of frequency maps which minimizes the lensing noise is the standard ILC, for all the estimators we consider.
Specifically, we show the following:
if the spatial weights $F^\kappa_{\bm{\ell},\bm{L}-\bm{\ell}}$ of the estimator $\hat{\kappa}$ depend on the frequency weights $\bm{w}_{\ell}$ only through the map noise $C^\text{tot}_\ell$, then the frequency weights $\bm{w}_{\ell}$ that extremize $\mathcal{N}_L[\bm{w}]$ are the ILC weights.
To minimize the noise subject to the constraint $\bm{w}^T_{\ell}\bm{1}=1$ we set 
\beq
\bal
\frac{\delta \mathcal{N}_L[\bm{w}]}{\delta \bm{w}_{\ell}}
&=
4 N_L^2 F^2_{\bm{\ell},\bm{L}-\bm{\ell}} \left[C^\text{tot}_\ell+C^\text{tot}_{|\bm{L}-\bm{\ell}|}\right]
\bm{C}_\ell \bm{w}_{\ell}+
4 N_L^2 \int \frac{d^2 \bm{\ell}'}{(2\pi)^2}
F_{\bm{\ell}',\bm{L}-\bm{\ell}'}
\frac{\delta F_{\bm{\ell}',\bm{L}-\bm{\ell}'}}{\delta \bm{w}_{\ell}}
C^\text{tot}_{\ell'}C^\text{tot}_{|\bm{L}-\bm{\ell}'|}\\
&-
2 N_L \mathcal{N}_L[\bm{w}] 
\int \frac{d^2 \bm{\ell}'}{(2\pi)^2}
\frac{\delta F_{\bm{\ell}',\bm{L}-\bm{\ell}'}}{\delta \bm{w}_{\ell}}
f^\kappa_{\bm{\ell}',\bm{L}-\bm{\ell}'}\\
&=\lambda_{\bm{\ell},\bm{L}} \bm{1},
\eal
\eeq
where $\lambda_{\bm{\ell},\bm{L}}$ is some Lagrange multiplier. If the spatial weights $F^\kappa_{\bm{\ell},\bm{L}-\bm{\ell}}$ only depend on frequency weights $\bm{w}_{\ell}$ through $C^\text{tot}_\ell$ and $C^\text{tot}_{|\bm{L}-\bm{\ell}|}$, then 
\beq
\bal
\frac{\delta F^\kappa_{\bm{\ell}',\bm{L}-\bm{\ell}'}}{\delta \bm{w}_{\ell}}
&=
\frac{\partial F^\kappa_{\bm{\ell}',\bm{L}-\bm{\ell}'}}{\partial C^\text{tot}_{\ell'}}
\frac{\delta C^\text{tot}_{\ell'}}{\delta \bm{w}_{\ell}}
+
\frac{\partial F^\kappa_{\bm{\ell}',\bm{L}-\bm{\ell}'}}{\partial C^\text{tot}_{|\bm{L}-\bm{\ell}'|}}
\frac{\delta C^\text{tot}_{|\bm{L}-\bm{\ell}'|}}{\delta \bm{w}_{\ell}}\\
&= 
(2\pi)^2 \frac{\partial F^\kappa_{\bm{\ell},\bm{L}-\bm{\ell}}}{\partial C^\text{tot}_{\ell}}
\left[
\delta^D_{\bm{\ell}-\bm{\ell}'}
+
\delta^D_{\bm{\ell}-\bm{L}+\bm{\ell}'}
\right]
\bm{C}_\ell \bm{w}_{\ell}\\
&\equiv 
\mathcal{A}_{\bm{\ell},\bm{\ell}',\bm{L}}
\bm{C}_\ell \bm{w}_{\ell}
\eal
\eeq
where $\mathcal{A}_{\bm{\ell},\bm{\ell}',\bm{L}}$ is some scalar function. Plugging this in to our equation for $\delta \mathcal{N}_L[\bm{w}]/\delta \bm{w}_{\ell}$ gives
\beq
\frac{\delta \mathcal{N}_L[\bm{w}]}{\delta \bm{w}_{\ell}}
=
\mathcal{B}_{\bm{\ell},\bm{L}} 
\bm{C}_\ell \bm{w}_{\ell}
=
\lambda_{\bm{\ell},\bm{L}} 
\bm{1}
\eeq
for some scalar function $\mathcal{B}_{\bm{\ell},\bm{L}}$. Solving this equation for $\bm{w}_{\ell}$ gives $\bm{w}_{\ell} = \lambda_{\bm{\ell},\bm{L}}  \bm{C}^{-1}_\ell \bm{1}/\mathcal{B}_{\bm{\ell},\bm{L}}$. The Lagrange multiplier is chosen so the weights sum to unity, which yields the standard ILC weights, given by Eq.~\eqref{eq: ILC_weights}.

Note that the standard QE, bias-hardened and shear estimators have respective weights:
\beq
F^{\kappa}_{\bm{\ell},\bm{L}-\bm{\ell}}
=
\frac{f^\kappa_{\bm{\ell},\bm{L}-\bm{\ell}}}{2C^\text{tot}_\ell C^\text{tot}_{|\bm{L}-\bm{\ell}|}}
\quad
,
\quad
F^{\kappa^\text{BH}}_{\bm{\ell},\bm{L}-\bm{\ell}}
=
\frac{f^\kappa_{\bm{\ell},\bm{L}-\bm{\ell}}-N^\kappa_L \mathcal{R}_L f^s_{\bm{\ell},\bm{L}-\bm{\ell}}}{2C^\text{tot}_\ell C^\text{tot}_{|\bm{L}-\bm{\ell}|}}
\quad
,
\quad
F^{\text{shear}}_{\bm{\ell},\bm{L}-\bm{\ell}}
=
\cos(2 \theta_{\bm{L},\bm{\ell}-\bm{L}/2} ) 
\frac{C^0_\ell}{2 (C^\text{tot}_\ell)^2}
\frac{d \ln C^0_\ell}{d \ln \ell},
\eeq
which only depend on $\bm{w}_\ell$ through $C^\text{tot}_\ell$. Thus their noises are minimized by the ILC weights.

\section{Attempts at numerically minimizing the loss functions: approximations}
\label{app:failed_approximations}

Finding the temperature weights which minimize the desired combination of lensing noise and bias is a difficult problem.
Indeed, the space of parameters to optimize is large, since the temperature weights are free functions of scale and frequency.
For each choice of temperature weights, one needs to perform multiple lensing reconstructions on the foreground simulations to estimate the primary, secondary and trispectrum biases.
This step is computationally expensive.
This needs to be repeated for each choice of estimator (QE, shear, PSH, PH), and at each step when exploring the space of temperature weights.
Below, we describe approximations which dramatically simplify the problem, and point out their inaccuracies.

One may assume that each foreground is determined by a single spatial template, whose amplitude is rescaled at each frequency (Approximation 1).
This approximation is exact for the CMB and for the tSZ, ignoring relativistic corrections.
It is also excellent for the CIB, whose maps at different frequencies are highly correlated \cite{2017MNRAS.466..286M, 2011A&A...536A..18P}.

To simplify further, we note that most of the CMB lensing signal to noise comes from temperature multipoles around $\ell=3000$.
As a result, most of the foreground biases also comes from these multipoles.
At these multipoles, the ILC weights are slowly varying.
We may therefore neglect the scale dependence of the ILC weights $\bm{w}^\text{ILC}$, and look for scale-independent optimal weights $\bm{w}$ (Approximation 2).
This approximation dramatically reduces the dimensionality of the minimization problem, making it much more tractable.
In this approximation, the primary $\mathcal{P}^s_{0,L}$, secondary $\mathcal{S}^s_{0,L}$ and trispectrum $\mathcal{T}^s_{0,L}$ biases to $A_\text{lens}$ only need to be computed at one frequency (e.g., 150~GHz).
They can then simply be rescaled to any other frequency.
Furthermore, because the temperature weights are scale-independent, the bispectrum and trispectrum biases only need to be computed from simulations once. 
They can then be rescaled for any new choice of temperature weights.
In short, approximations 1 and 2 allow us to compute the foreground biases from simulation only once, rather than having to do it for every frequency and for every choice of weights $\bm{w}_\ell$.

Finally, we expect the optimal multifrequency combination will be close to the standard ILC.
We indeed show this to be true for the profile hardened estimator when exploring part of the parameter space, going from ILC to some form of deprojection.
This motivates us to Taylor expand the noise about the ILC weights to second order in $\bm{\epsilon}$ (Approximation 3):
$\bm{w} = \bm{w}^\text{ILC}_{\ell=3000}+\bm{\epsilon}$.
This approximation dramatically speeds up the noise computation during the optimization step, and accurately predicts the variance and bias of $A_\text{lens}$ except at 27~GHz. A simple method for getting around this is to set the 27~GHz weight to zero, in which case the Taylor expansion agrees with the true noise to the subpercent level for the $\bm{\epsilon}$'s relevant for achieving bias$/$noise $\sim1/2$.

In this simplified case, the scaling of the foreground biases with the frequency weights is simply:
\beq
\text{bias}(C^\kappa_L)[\bm{w}]
=
\sum_s
\left[\mathcal{P}^s_{0,L} + \mathcal{S}^s_{0,L} \right]
\left(\frac{\bm{A}_s\cdot\bm{w}}{A_{s,150}}\right)^2+
\sum_s 
\mathcal{T}^s_{0,L}
\left(\frac{\bm{A}_s\cdot\bm{w}}{A_{s,150}}\right)^4
\eeq
where $A_{s,150}$ is the scale-factor for the 150 GHz channel. 
In practice, this expression can be easily generalized to include the lensing bias from correlated foregrounds, e.g., CIB and tSZ.
In summary, we are thus looking for a handful of parameters $\left( \epsilon_\nu \right)_{\nu=1...5}$, which minimize the sum of squared variance and squared bias on $A_\text{lens}$.

However, we found approximations 2 and 3 not to be sufficiently accurate.
Assuming the weights to be scale-independent leads to a significant misestimation of the variance and bias of $A_\text{lens}$, at the $\mathcal{O}(1)$ level.

\begin{figure*}[!h]
    \centering
    \includegraphics[width=0.48\linewidth]{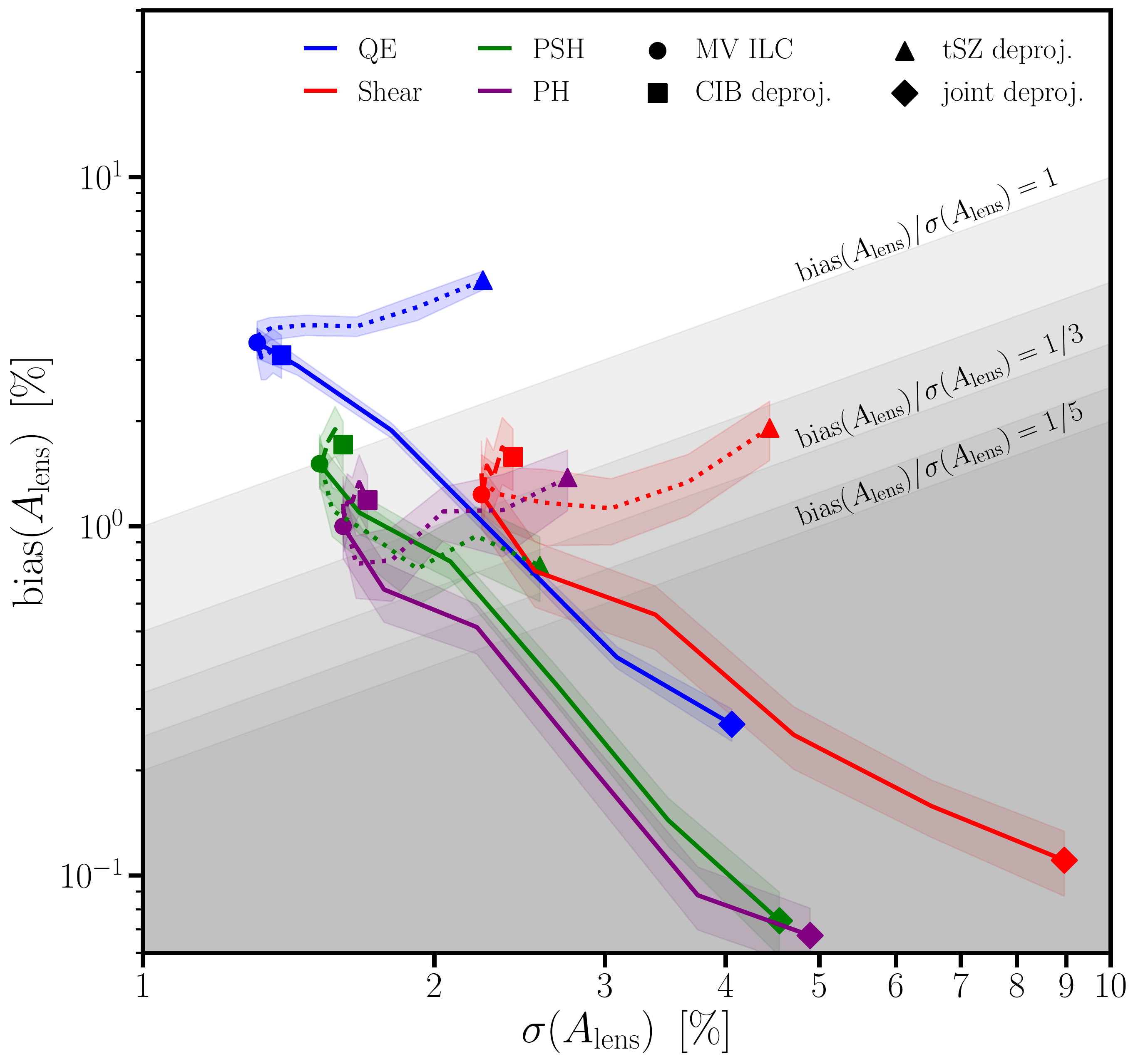}
    \includegraphics[width=0.48\linewidth]{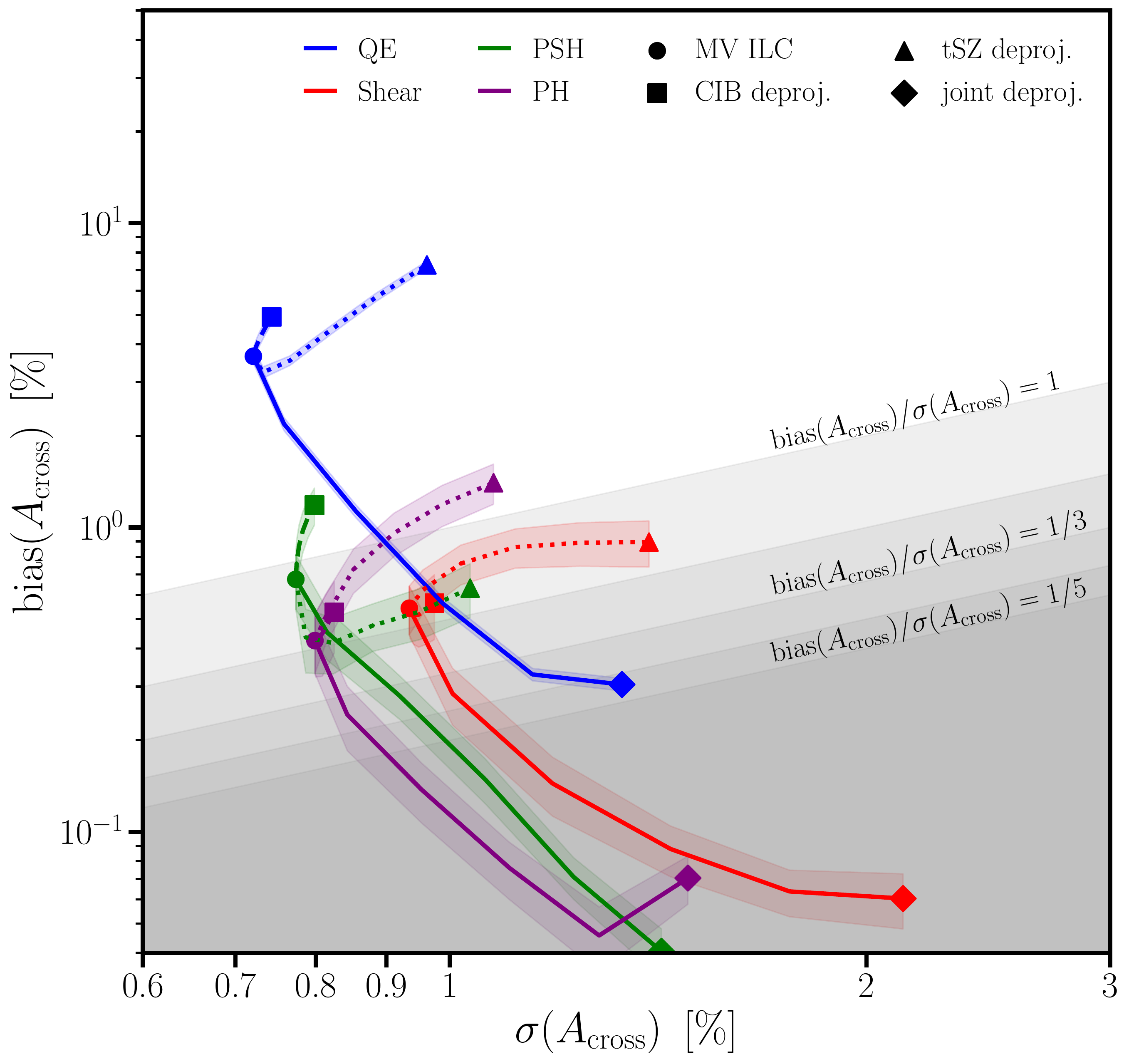}
    \caption{The same as Fig.~\ref{fig:money_plot}, using the more agressive masking technique described in App. \ref{app:mask}. When calculating $\sigma$, we neglect any differences in sky coverage due to more of the sky being masked (8\% of the sky as opposed to 3\% with the fiducial mask used in the main text), as this is a negligible effect.}
    \label{fig:fancier_mask}
\end{figure*}

\section{Dependence of the lensing biases on the masking}
\label{app:mask}

We explore a more aggressive masking approach by running the matched filter at all the SO frequencies, masking only the sources (using disks with a 2 arcmin radius) that are detected with $5\sigma$ significance, and using the union of the masks at each frequency.  
This allows us to find much more IR (radio) sources, which are brighter at higher (lower) frequencies.
In an analysis on real data, this may make it more delicate to estimate the normalization and mean field of the lensing estimators \cite{LemboInPrep}, which has not been taken into account in our lensing biases calculations.

We note that in practice, masking changes the observed map covariance $\bm{C}_\ell$, which in turn will influence the form of the ILC weights. For simplicity we have ignored this effect, and thus use the same ILC weights for the more aggressive masking approach as we did in the main text, which are derived using the theory foreground spectra from \cite{2013JCAP...07..025D}.

The resulting biases from this masking technique are shown in Fig.~\ref{fig:fancier_mask}. In cross-correlation, the aggressive masking reduces the foreground bias for all estimators, as expected. In particular, at ILC ($t=0$) the biases to the cross-correlation for PH, Shear, PSH, and QE are reduced by $\sim30\%$, $40\%$, $10\%$, and $60\%$  respectively, relative to the biases calculated using the single mask at 150 GHz. In auto-correlation, the biases for PH, Shear and PSH are similarly reduced using the more aggressive masking technique. However, the bias to the QE at ILC increases by $50\%$. This can be understood as follows. In auto-correlation, both the primary and trispectrum terms contribute to the foreground bias.
The aggressive masking reduces both, but reduces the trispectrum much more.
For the QE, a ``lucky'' cancellation allows the dominant negative primary bias (at low lensing-$L$) to cancel the dominant positive trispectrum bias (at high lensing-$L$).
By reducing the trispectrum term more than the primary term, the aggressive masking spoils this cancellation, thus enhancing the QE bias in auto-correlation.

\section{Dependence of the lensing biases on $\ell_{\text{max,}T}$}
\label{app:lmaxt}

\begin{figure*}[!h]
    \centering
    \includegraphics[width=0.48\linewidth]{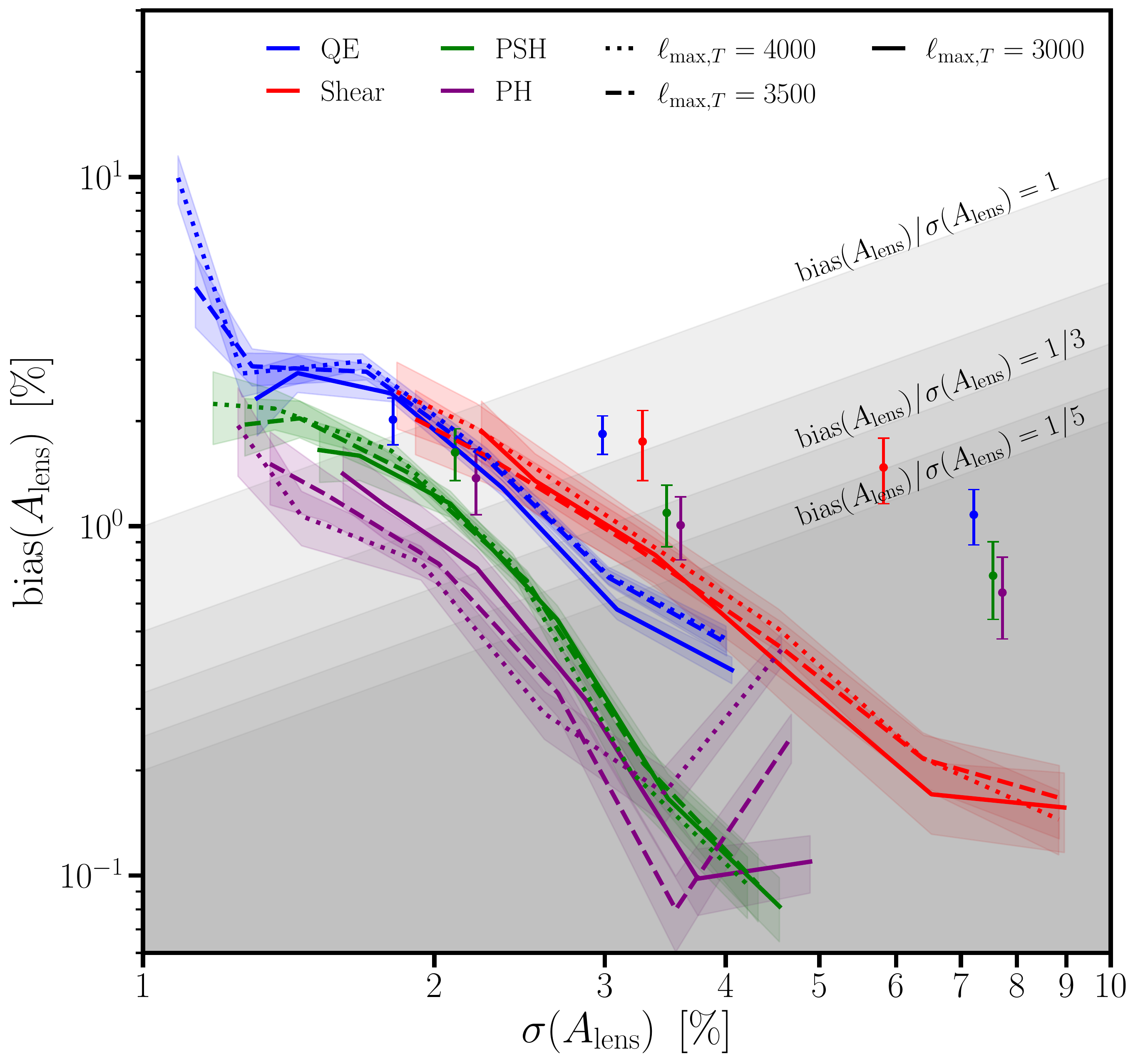}
    \includegraphics[width=0.48\linewidth]{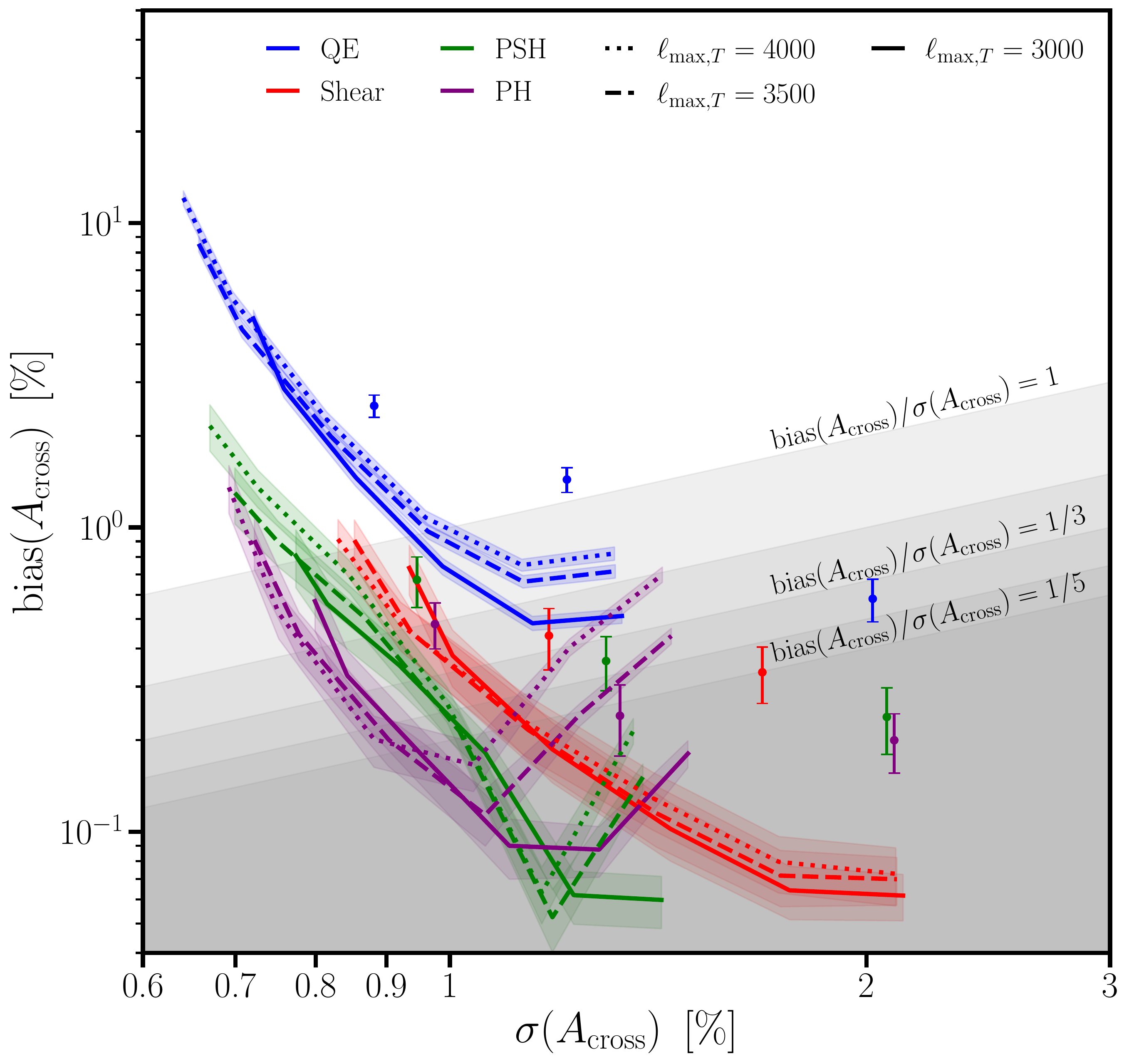}
    \caption{Noise vs bias for several $\ell_{\text{max},T}$'s. For $\ell_{\text{max},T} = 3000,\,3500,$ and $4000$ we plot the line going from the standard ILC to joint deprojection, with the linestyle denoting the $\ell_{\text{max},T}$. For $\ell_{\text{max},T} = 1500,\,2000,$ and $2500$ we plot the noise and bias at the standard ILC for each estimator (denoted by the circles). Note that for $\ell_{\text{max},T} = 1500$, the Shear estimator's noise is larger than 10\% (3\%) for the auto- (cross-) correlation, extending beyond the domain of the plot.}
    \label{fig:varied_lmaxT}
\end{figure*}

In this section we explore how our results change with $\ell_{\text{max},T}$. As shown in Fig.~\ref{fig:varied_lmaxT}, we still find that profile hardening outperforms the remaining estimators for any $\text{bias}/\text{noise}$ cutoff and for any $\ell_{\text{max},T}$. We also find that small improvements are possible by pushing to a higher $\ell_{\text{max},T}$ with more joint deprojection. For the auto-correlation, the bias to the profile hardened estimator can be reduced below half of the statistical uncertainty at a $30\%$ noise penalty by choosing $\ell_{\text{max},T}=4000$ and $t\sim0.4$ (as opposed to a $50\%$ penalty with $\ell_{\text{max},T}=3000$ and $t\sim0.2$), relative to the standard QE at ILC with $\ell_{\text{max},T}=3000$. For the cross-correlation, the noise penalty for achieving $\text{bias}/\text{noise}<1/2$ with the profile hardened estimator can be reduced from $15\%$ to $10\%$ by similarly increasing $\ell_{\text{max},T}$ and $t$.

We note that for sufficiently high $t$, the biases and noises become roughly insensitive to the value of $\ell_{\text{max},T}$. This is due to the map noise blowing up at high-$\ell$ as the weights are pushed to joint deprojection, as shown in Fig.~\ref{fig:various_ILC_noises}. This in turn down-weights high-$\ell$ modes and acts as an effective cutoff with $\ell_\text{max eff}<\ell_{\text{max},T}$. Instead of partial joint deprojection (Eq.~\eqref{eq:walks}), one could naively use the standard ILC weights and reduce $\ell_{\text{max},T}$ to remove these high-$\ell$ modes, and hence the bias, in a potentially a more optimal way. However, in practice we find it more optimal to use a high $\ell_{\text{max},T}$ with joint deprojection than a lower $\ell_{\text{max},T}$ with the standard ILC weights. This is illustrated in Fig.~\ref{fig:varied_lmaxT}, where we plot the bias and noise at ILC for $\ell_{\text{max},T} = 1500,\,2000,$ and $2500$ (individual circles). While lowering the $\ell_{\text{max},T}$ decreases the bias (as expected), the noise cost for doing so is much larger than increasing $t$ with a higher  $\ell_{\text{max},T}$ (as illustrated by the three lines for each estimator).

\end{document}